\documentclass{iopart}

\usepackage{graphicx}
\usepackage{amssymb}
\usepackage{bm}
\newcommand{\pdag}{{\phantom{\dagger}}}

\begin{document}

\title{Unitary Perturbation Theory Approach to Real-Time Evolution  
Problems}

\author{A Hackl$^1$ and S Kehrein$^2$}

\address{$^1$Institut f\"ur theoretische Physik, Universit\"at zu K\"oln, Z\"ulpicher Str. 77, 50937 K\"oln, Germany}
\address{$^2$Arnold Sommerfeld Center for Theoretical Physics and  
CeNS, Department f\"ur Physik, Ludwig-Maximilians-Universit\"at, Theresienstrasse 37, 80333 M\"unchen, Germany}

\ead{ah@thp.uni-koeln.de}
\begin{abstract}
We discuss a new analytical approach to real-time evolution in  
quantum many-body systems.
Our approach extends the framework of continuous unitary  
transformations such that it amounts
to a novel solution method for the Heisenberg equations of motion for  
an operator. It is our purpose
to illustrate the accuracy of this approach by studying dissipative  
quantum systems on all time scales.
In particular, we obtain results for non-equilibrium correlation  
functions for general initial conditions.
We illustrate our ideas for the exactly solvable dissipative  
oscillator, and, as a non-trivial model, for
the dissipative two-state system.
\end{abstract}
\maketitle
\section{\label{intro}Introduction}

Strongly correlated many-body systems are challenging due to their  
highly non-trivial interplay between
different energy scales. In experiment, many of their properties are  
probed by
measuring their linear response to weak external driving forces. An  
additional branch of
physical phenomena is currently being explored by driving quantum  
many-body systems far out of equilibrium
and studying such new states of matter. In particular,
recent progress in atomic physics has made it possible to tune  
systems of
ultracold atoms at will between different interacting regimes. For  
example, a seminal experiment
by Greiner et al.  \cite{Greiner2002}
shows collapse and revival phenomena in atomic Bose gases after an  
interaction quench,
where excellent isolation from the environment allows the observation  
of such non-equilibrium
behavior at remarkably long time scales.

 From the theoretical side, non-equilibrium quantum systems are very  
challenging since
many well-established methods from equilibrium physics
cannot be directly applied. Significant progress has been achieved
in low-dimensional systems and for quantum impurity systems with  
numerical
methods like NRG and DMRG: Their newly developed extensions
called time-dependent density matrix renormalization group (TD-DMRG) 
\cite{Schollwoeck_TDDMRG}
and time-dependent numerical renormalization group (TD-NRG)\cite 
{Anders_KM,Anders_SB}, were, e.g.,
successfully applied to interaction quenches in the Bose-Hubbard  
model \cite{Kollath}
and to impurity spins coupled to bosonic or fermionic baths \cite 
{Anders_SB}.
More recent applications of these methods also include interaction  
quenches in fermionic
lattice systems in one dimension \cite{Manmana2007} and the  
calculation of steady state
currents through nano-devices \cite{Anders2008}.

The extension of conventional analytical techniques seems to be
more difficult. For example, the renormalization group approach,  
which is
designed to construct effective low-energy  Hamiltonians, is not  
ideally suited
to understand non-equilibrium situations, where energy scales well above
the low-energy sector can be excited.
The situation is not much better in weakly interacting systems, where
diagrammatic expansions are usually the first tool to analyze
ground states and their elementary excitations, and often yield  
reliable results in
thermal equilibrium.
Nevertheless, even in weakly interacting systems perturbation theory  
is often not
capable of describing the long-time behaviour
of non-equilibrium observables, since truncation errors can become  
uncontrolled
at large time scales.

In this paper, we present a new analytical approach to the real-time  
evolution problem which
merges the advantages of perturbation theory and renormalization  
group theory, and at the same time,
leaves behind their shortcomings mentioned above. In two short  
previous publications,
this approach has already been introduced and was applied to two  
different problems. \cite{Hackl2007,Moeckel2008}
Here we give a somehow more pedagogical introduction, provide more  
details and discuss its general applicability to the
field of dissipative quantum systems. In the last twenty years,  
quantum physics in a dissipative
environment has played an important role in solid state
physics, quantum optics, quantum computing, chemical and even biological
systems (for a review, see Refs.~\cite{Leggett1987,Weiss1999}).
In addition, this field is especially
suitable for checking the accuracy
of our approximation scheme since many results from other approaches  
are known.
However, our approach is applicable to
a much wider class of problems, including also lattice models,
see for example Ref.~\cite{Moeckel2008}.

\subsection{Summary of results}

Our approach is based on an analogy to canonical perturbation theory  
in classical mechanics.
We give a simple illustration of canonical perturbation theory
and show how canonical transformations can improve perturbative  
expansions in
real-time evolution problems. We show that an analogous  
implementation for quantum
many-body systems is possible, based on Wegner's flow equation  
approach \cite{Wegner1994}. Independently, the same approach 
has been developed in the field of high energy physics \cite{Glazek1993}.
Using this approach, we reproduce the exact solution for a quantum  
dissipative
oscillator and show that efficient and precise numerical solutions of  
the analytical equations
can be obtained. We also illustrate the failure of naive perturbation  
theory in this simple
quantum mechanical system.

For the spin-boson model, we obtain results that are in excellent  
agreement with known
results. In the regime of weak coupling to the bath, we reproduce  
exact results for
the decoherence of a spin without tunnel splitting. For finite tunnel  
splitting, we calculate
the non-equilibrium correlation function of the spin projection and  
also obtain
the quantitatively correct coherent decay of the spin polarization in  
the weakly damped Ohmic
regime.

\subsection{Outline}

In section \ref{feq} we give an introduction into the basic ideas of  
our method
and its technical details. First we motivate our approach using an  
analogy
from classical mechanics. Then we discuss the method in detail. In  
section \ref{dho},
we apply this approach to a simple exactly solvable model which is  
useful to understand
the technical details of our method. In section \ref{tss}, we apply  
our method to the non-trivial
spin-boson model to analyze the reliability of our approximation scheme.
A brief summary and outlook concludes the paper in section \ref{disc}.

\section{\label{feq}Real-time evolution with the flow-equation method}

\subsection{Motivation: Canonical perturbation theory in classical  
mechanics}

There exist well-established methods to handle time-dependent
perturbations in classical mechanics, and early attempts to handle  
these problems date back already to Newton, \cite 
{Textbook_analyticalmechanics} who considered
the small distortion of the moon orbit caused by the gravitational  
force of the sun.

Much progress in this field has been motivated by the ever increasing  
accuracy
of observational data for planetary motion and satellites, and the  
need to make accurate
predictions based on this. The common approaches to these problems  
are collectively
summarized as ``canonical perturbation theory".\cite 
{Textbook_analyticalmechanics}
The basic idea behind canonical perturbation theory can be simply  
illustrated by the Hamilton
function of a weakly perturbed classical harmonic oscillator,

\begin{equation}
H=\frac{1}{2}p^2+ \frac{1}{2}q^2 +\frac{g}{4}q^4,
\label{hamilton}
\end{equation}

where a quartic anharmonic term with a small coupling $g \ll 1$  
perturbs the trajectory.
We consider the initial conditions $q(t=0)=0$ and $p(t=0)=v+\frac{3}{8}gv^3$ in the following.

By exploiting the smallness of the quartic perturbation, one might be  
tempted to employ naive
perturbation theory which uses a series in powers of $g$ as an ansatz  
for the perturbed solution
$q(t)$,

\begin{equation}
q(t)=q^{(0)}(t)+g q^{(1)}(t)+ O(g^2).
\label{naive}
\end{equation}

The trajectory $q^{(0)}(t)$ is just the solution of the unperturbed  
problem,

\begin{equation}
q^{(0)}(t)=v\sin(t).
\end{equation}

 From Hamilton's equations,

\begin{eqnarray}
\frac{\partial H}{\partial p} &=& \dot{q} \nonumber\\
\frac{\partial H}{\partial q} &=& -\dot{p},
\end{eqnarray}

we obtain the equation of motion

\begin{equation}
\ddot{q}^{(1)}(t)= -q^{(1)}(t) - v^3 \sin^3(t),
\end{equation}

which has the solution

\begin{equation}
q^{(1)}(t)=  \frac{3}{8}v^3 \sin(t)-\frac{v^3}{8} \bigl( \sin(t)  
\cos^2(t) + 2 \sin(t) -3t \cos(t)\bigr).
\label{naiveresult}
\end{equation}

This result already reveals the caveat of this approach, since the so  
called secular term $3t\cos(t)$
yields an error growing unbounded in time. In fact, it is a well  
known general result from classical
mechanics that such secular terms invalidate naive perturbation theory
for large times.

Naive perturbation theory can be much improved in the framework of  
canonical perturbation theory.
This approach first transforms the Hamilton function to a suitable  
normal form, and
after solving the equations of motion in this canonical basis, the  
normal coordinates are
reexpressed through the old coordinates. In this manner secular terms  
can be avoided.
To implement this idea, we first look for a canonical transformation  
of variables

\begin{eqnarray}
(q,p) \rightarrow (Q,P)
\end{eqnarray}

that brings the Hamilton function to the following normal form,  
denoted by $\tilde{H}$,

\begin{equation}
\tilde{H}=H_0+g\alpha H_0^2+O(g^2) \quad \textrm{with} \quad H_0=\frac 
{1}{2}P^2+\frac{1}{2}Q^2.
\label{normal}
\end{equation}

\begin{figure}
\includegraphics[clip=true,width=8.0cm]{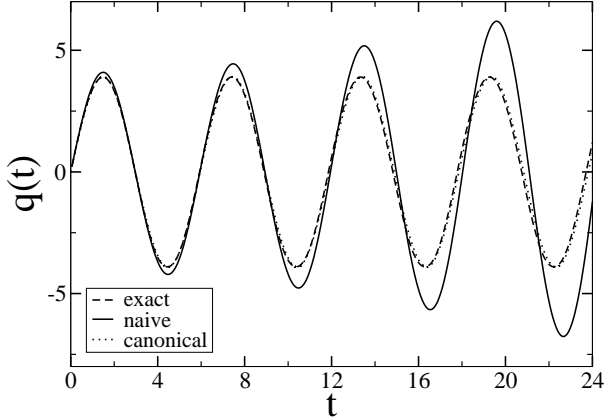}
\caption{\label{canonicalpic} We compare the different approaches to  
solve the equations of motion
for the anharmonic oscillator from Eq. (\ref{hamilton}). The  
difference between the numerically exact solution
and canonical perturbation theory according to Eq. (\ref 
{canonicalresult}) can hardly be noticed. Naive perturbation theory
yields large errors already after a few oscillations, with an error  
that grows linear in time $t$.
Our parameters are $v=4$ and coupling strength $g=0.01$.
}
\end{figure}

It is easy to see that the Poisson bracket $\{H_0,H_0^2\}$ vanishes,  
hence the equations of motion for $Q$ and $P$
can be solved trivially. These variables obey the initial conditions $Q(0)=0$ and  
$P(0)=0$. In our example, the corresponding  
transformation of variables is

\begin{eqnarray}
q(t)&=&Q(t)-\frac{3}{32}g \biggl( 3 P^2(t) Q(t) +\frac{5}{3}Q^3(t) 
\biggr) \nonumber\\
&+& O(g^2), \nonumber\\
p(t)&=&P(t)+ \frac{3}{32}g\biggl( 5P(t)Q^2(t) +P^3(t) \biggr)  
\nonumber\\
&+& O(g^2)
\end{eqnarray}

which is performed perturbatively in the parameter $g$.  The normal  
form (\ref{normal}) has
been chosen such that the equation of motion for the new variables
$P(t)$ and $Q(t)$ can now be solved exactly,
without producing any secular term. Using this strategy, the final  
result is

\begin{eqnarray}
q(t)&=& v \sin(\omega t)\nonumber\\
&-&\frac{3}{32}gv^3 \biggl( 3\cos^2(\omega t) \sin(\omega t) +\frac{5} 
{3} \sin^3(\omega t) \biggr) \nonumber\\
  &+& O(g^2), \\
  p(t)&=& v\cos(\omega t) +\frac{3}{32}gv^3\biggl( 5\sin^2(\omega t)  
\cos(\omega t) + \cos^3(\omega t)\biggr)\nonumber\\
  &+& O(g^2)
\label{canonicalresult}
\end{eqnarray}

where $\omega=1+\frac{3}{4}g E_0$ and $E_0=p(0)^2/2$. We show a  
comparison
of naive perturbation theory and canonical perturbation theory in  
Fig. \ref{canonicalpic}, which demonstrates
the usefulness of canonical perturbation theory.

Canonical perturbation theory yields renormalized parameters of the  
unperturbed problem, but does not
lead to secular terms. This improvement becomes directly visible by  
expanding the contribution $\sin(\omega t)$
in powers of~$g$,

\begin{eqnarray}
\sin(\omega t) &=& \sin\biggl((1+ \frac{3}{4}gE_0)t\biggr)\nonumber\\
  &=&\sin(t) +\frac{3}{4}gE_0t \cos(t) +O(g^2).
\end{eqnarray}

This expansion generates the secular term from Eq. (\ref 
{naiveresult}) occuring in naive perturbation theory,
making it obvious that canonical perturbation theory contains a  
summation over secular terms
which in total yield a much improved result in comparison to naive  
perturbation theory.
Notice that this is true in spite of the fact that {\em both  
approaches have been expanded to the same power
in the small coupling constant}.\\

\subsection{Perturbation theory in non-equilibrium quantum mechanics}

In quantum many-body systems, the canonical way to evaluate the real  
time evolution of observables
starting from some non-thermal initial state
is the Keldysh technique \cite{Mahan}, which defines a contour  
ordered S-matrix in order
to develop perturbative expansions for non-equilibrium Greens functions.
Just as in our example of naive perturbation theory, secular terms  
can occur in any finite order of perturbation
theory, and there is no universal solution for how to sum up these  
secular terms.
These difficulties can make it very difficult or even impossible to  
study
the transient evolution of observables into a steady state.

One might wonder why an analogue of canonical perturbation theory
for quantum many-body systems had not been developed earlier. The  
basic reason is the notorious difficulty
to transform quantum-many body systems into normal form. In addition,
the continuum of energy scales often causes often non-perturbative
effects in coupling constants. Since the advent of renormalization  
group techniques,
the much easier problem of constructing effective low-energy theories  
has attracted
most of the attention. The more general problem of constructing  
normal forms of interacting
Hamiltonians has been successfully treated only during the last years.

\subsection{Flow equation approach}

\begin{figure}
\includegraphics[clip=true,width=3.2cm,angle=90]{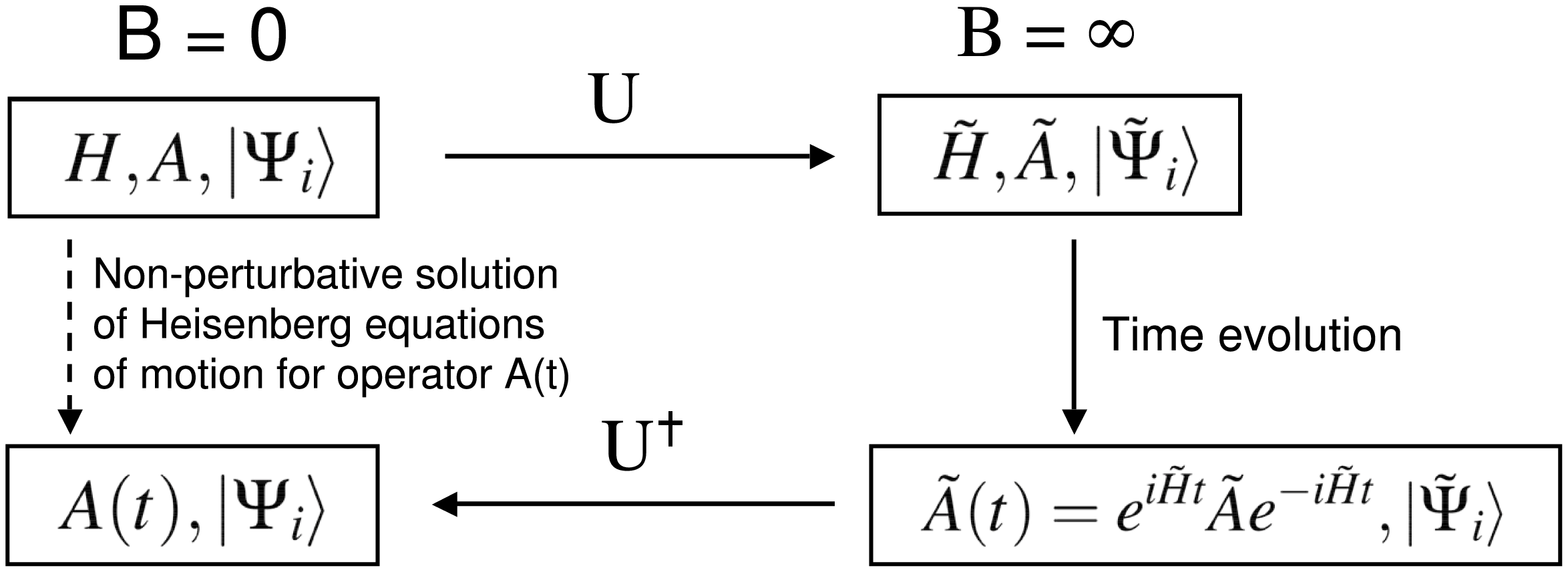}
\caption{\label{forwardback} How to make use of the flow-equation
method to implement the analogue of canonical perturbation theory in  
quantum mechanics.
U denotes the full unitary transformation
that relates the $B = 0$ to the $B = \infty$ basis.
}
\end{figure}

A general description how to transform interacting quantum many body  
systems
into a non-interacting normal form has been given in 1994 by F.  
Wegner.\cite{Wegner1994}
Let us briefly review the basic ideas of the flow equation approach
(for more details see Ref.\cite{Kehrein_STMP}). A many-body  
Hamiltonian~$H$
is diagonalized through a sequence of infinitesimal unitary  
transformations with
an anti-hermitean generator~$\eta(B)$,
\begin{equation}
\frac{dH(B)}{dB}=[\eta(B),H(B)] \ ,
\label{eqdHdB}
\end{equation}
with $H(B=0)$ the initial Hamiltonian.
The ``canonical" generator \cite{Wegner1994} is the commutator of the
diagonal part~$H_{0}$ with the interaction part~$H_{\rm int}$ of the  
Hamiltonian,
$\eta(B)\stackrel{\rm def}{=}[H_{0}(B),H_{\rm int}(B)]$. It can be  
formally shown
that the choice of the canonical generator is by construction  
suitable to eliminate
interaction matrix elements with energy transfer $\Delta E=O\bigl(1 /  
\sqrt{B}\bigl)$.
Under rather general conditions, an increasingly energy diagonal  
Hamiltonian is obtained.
For $B=\infty$ the Hamiltonian will be energy-diagonal and we denote  
parameters and operators in
this basis by~$\tilde{~}$, e.g.\ $\tilde H=H(B=\infty)$.

Usually, this scheme cannot be implemented exactly. The generation of  
higher and higher order interaction
terms in (\ref{eqdHdB}) makes it necessary to truncate the scheme in  
some order of a suitable systematic
expansion parameter (usually the running coupling constant).
Still, the infinitesimal nature of the approach
makes it possible to deal with a continuum of energy
scales and to describe non-perturbative effects. This had
led to numerous applications of the flow equation method
where one utilizes the fact that the Hilbert space is not
truncated as opposed to conventional scaling methods.

In Ref. \cite{Hackl2007}, these features have been exploited in  
order to
develop an analogue of canonical perturbation theory in classical  
mechanics for
quantum many-body systems.
The general setup is described by the diagram in Fig. \ref 
{forwardback}, where $| \Psi_i \rangle$ is some
initial non-thermal state whose time evolution
one is interested in. However, instead of following its full
time evolution it is usually more convenient to study the real
time evolution of a given observable A that one is interested in.  
This is done by
transforming the observable into the diagonal basis in
Fig. \ref{forwardback} ({\sl forward transformation}):

\begin{equation}
\frac{dO}{dB}=[\eta(B),O(B)]
\label{operatorflow}
\end{equation}
with the initial condition $O(B = 0) = A$. The key
observation is that one can now solve the real time
evolution with respect to the energy-diagonal $\tilde H$ exactly,
thereby avoiding any errors that grow proportional to
time (i.e., secular terms): this yields $\tilde A(t)$. Now since
the initial quantum state is given in the $B=0$ basis,
one undoes the basis change by integrating (\ref{operatorflow}) from
$B = \infty$ to $B = 0$ ({\sl backward transformation}) with the
initial condition $O(B = \infty) = \tilde A(t)$. One therefore
effectively generates a new non-perturbative scheme for solving
the Heisenberg equations of motion for an operator,
$A(t) = e^{iHt} A(0) e^{-iHt}$, in exact analogy to canonical  
perturbation theory.

\section{\label{dho}The dissipative harmonic oscillator}

The dissipative harmonic oscillator is a widely used toy model in the  
field of quantum optics and
is also used in many other contexts.\cite{Decker1981} It describes a  
quantum oscillator of frequency $\Delta$ coupled linearly to a heat
bath consisting of bosonic normal modes $b_k$.

\begin{eqnarray}
H &=& \Delta b^\dagger b^\pdag +\sum_k \omega_k b_k^\dagger b_k^\pdag  
+E_0 \nonumber\\
   &+& \sum_k \lambda_k (b+b^\dagger ) (b_k^\pdag +b_k^\dagger )
\label{dhomodel}
\end{eqnarray}

The operators $b_k$ fulfill canonical commutation relations

\begin{equation}
[b_k^\pdag,b_{k^\prime}^\dagger ]= \delta_{kk^\prime}
\end{equation}

and their influence on physical properties of the quantum oscillator  
can be fully described by the spectral function

\begin{equation}
J(\omega) \stackrel{\rm def}{=} \sum_k \lambda_k^2 \delta (\omega - 
\omega_k )
\label{spectrum}
\end{equation}

In experiment,
the high frequency part of $J(\omega)$ only affects the short time- 
response of the system, and thus, this part of the spectrum
is usually cut off from $J(\omega)$. In consequence, the function $J 
(\omega)$ is commonly approximated by a power-law behavior
$J(\omega)\propto \omega^s, \omega < \omega_c$. Three different  
regimes of the exponent are distinguished,
for $0 < s < 1$ the bath is called ``Subohmic", for $s=1$ the bath is  
called ``Ohmic" and for $s > 1$ ``Superohmic".

We imagine that the system is prepared in a well-defined quantum  
state at some time $t_0$ and the subsequent real time evolution
of physical quantities is then defined by the Hamiltonian (\ref 
{dhomodel}). The flow equation method
does not restrict the class of possible initial states, but in this  
work we will
consider the system-bath complex at time $t=0$ to be prepared in a  
product state

\begin{equation}
| a \rangle \otimes | \Omega \rangle
\label{coherentstate}
\end{equation}

with the quantum oscillator in a coherent state

\begin{equation}
| \alpha\rangle=e^{-\frac{a^2}{2}} \sum_{n=0}^{N}\frac{a^n}{\sqrt 
{n!}}| n\rangle, ~ a \in \mathbb{R}
\end{equation}

and the heat bath in the bosonic vacuum state $| \Omega \rangle$. In  
such a state,
the displacement $\langle \hat{x} \rangle=1/\sqrt{2}\langle(b+b^ 
\dagger)\rangle$ will be finite and the
effects of decoherence and dissipation will manifest themselves in  
the real time evolution
of the observable $\langle \hat{x} (t)\rangle$. The flow equation  
method, outlined in section \ref{feq}, can solve
this problem exactly without approximations. First, the flow  
equations for the Hamiltonian are derived.
Then the real time evolution of the operators $b$ and $b^\dagger$ is  
implemented by their time-dependent flow equations.
This will allow us to study the forward-backward transformation  
scheme in the context of an exactly
solvable model.

\subsection{Diagonalization of the Hamiltonian}

As the first step of our program, the coupled form of the Hamiltonian  
(\ref{dhomodel}) will be diagonalized by infinitesimal unitary
transformations. Due to the quadratic nature of the Hamiltonian (\ref 
{dhomodel}), the flow equations for the Hamiltonian
can be derived in closed form, as shown in Ref. \cite 
{KehreinMielke}.

Commuting the interaction part with the non-interacting boson part of  
(\ref{dhomodel}) yields
the canonical generator
$\eta(B)=[\Delta b^\dagger b^\pdag + \sum_k \omega_k b_k^\dagger b_k^ 
\pdag ,\sum_k \lambda_k (b+b^\dagger ) (b_k^\pdag +b_k^\dagger )]$
of unitary transformations:

\begin{eqnarray}
\eta(B)&=&\sum_k \lambda_k(B)\Delta(B) (b^\dagger-b )(b_k+b_k^\dagger) 
\nonumber\\
&+&\sum_k\omega_k(B)\lambda_k(B)(b+b^\dagger)(b_k-b_k^\dagger)
\end{eqnarray}

This generator leads to additional terms in the flowing Hamiltonian  
that violate the form invariance of
equation (\ref{dhomodel}). The flowing Hamiltonian preserves its  
initial form if the generator is extended
by additional terms. In the following, we omit the explicit  
dependence of coefficients on the parameter $B$.

\begin{eqnarray}
\eta(B) &=& \sum_k \eta_k^{(1)}(b-b^\dagger )(b_k+b_k^\dagger) 
\nonumber\\
&+&\sum_{k,q}\omega_k \lambda_k (b+b^\dagger)
(b_k-b_k^\dagger) \nonumber\\
&+& \sum_{k,q} \eta_{k,q} (b_k + b_k^\dagger ) (b_q-b_q^\dagger ) +  
\eta_b (b^2-b^{\dagger 2})
\label{gendho}
\end{eqnarray}

with the coefficients

\begin{eqnarray}
\eta_k^{(1)}&=&-\lambda_k \Delta \tilde{f}(\omega_k,B) \nonumber\\
\eta_k^{(2)}&=& \lambda_k \omega_k \tilde{f}(\omega_k,B) \nonumber\\
\eta_{k,q} &=& -\frac{2\lambda_k \lambda_q \Delta \omega_q} 
{\omega_k^2-\omega_q^2}(\tilde{f}(\omega_k,B ) + \tilde{f} 
(\omega_q,B ))\nonumber\\
\eta_b &=& -\frac{1}{4\Delta} \frac{d\Delta}{dB}
\end{eqnarray}

which are chosen such that they leave the flowing Hamiltonian form  
invariant.
For this purpose the function $\tilde{f}(\omega_k,B)$
is still arbitrary, except for the obvious requirement $\lambda_k(B= 
\infty)=0$.
For our numerical evaluations of the flow equations later on, we will  
chose
$\tilde{f}(\omega_k,B)=-\frac{\omega_k-\Delta}{\omega_k+\Delta}$,
which leads to good convergence properties in the limit $B\rightarrow 
\infty$

The flowing parameters of the transformed Hamiltonian are governed by  
the coupled differential equations

\begin{eqnarray}
  \frac{d\Delta(B)}{dB} &=& 4 \sum_k \eta_k^{(2)}\lambda_k \nonumber\\
  \frac{dE_0(B)}{dB}    &=& 2\sum_k \eta_k^{(2)}\lambda_k +2\sum_k  
\eta_k^{(1)}\lambda_k \nonumber\\
  \frac{d\omega_k}{dB}  &=& O(\frac{1}{N}) \nonumber\\
  \frac{d\lambda_k}{dB} &=& \Delta\eta_k^{(1)} +\omega_k \eta_k^ 
{(2)}  \nonumber\\
  && + ~ 2\sum_q\eta_{k,q} \lambda_q + 2 \eta_b \lambda_k
\end{eqnarray}

The renormalization of the bath frequencies $\omega_k$ will have a  
vanishing effect on time-dependent observables in the
thermodynamic limit $N \rightarrow \infty$. Therefore, the flow  
equations of these energies can be ignored for our purposes.
In the limit $B \rightarrow \infty$, the Hamiltonian is diagonalized  
and the tunneling matrix element $\Delta$ is
renormalized to some value $\tilde{\Delta}$.

\begin{equation}
\tilde{H} = \tilde{\Delta} b^\dagger b + \sum_k \omega_k b_k^\dagger b_k
\label{hinfty}
\end{equation}

In order to proceed with the real-time evolution of observables, the  
second step of the program outlined in section \ref{feq}
requires the analogous transformation of the system operators.

The flow of the bosonic operator $b(B)$ is determined by the  
generator (\ref{gendho})
and yields the following structure

\begin{eqnarray}
b(B) &=& \beta (b+b^\dagger ) +\bar{\beta}(b-b^\dagger) \nonumber\\
&+& \sum_k \alpha_k(b_k+b_k^\dagger) + \bar{\alpha}_k(b_k-b_k^\dagger)
\label{bansatz}
\end{eqnarray}

with the flow equations

\begin{eqnarray}
\frac{d\beta(B)}{dB} &=& 2\eta_b \beta + 2\sum_k \alpha_k \eta_k^ 
{(2)} \nonumber\\
\frac{d\bar{\beta}(B)}{dB} &=& -2\eta_b \bar{\beta} -2 \sum_k \bar 
{\alpha_k} \eta_k^{(1)} \nonumber\\
\frac{d\alpha_k(B)}{dB} &=& 2\eta_k^{(1)} +2 \sum_q \eta_{k,q}  
\alpha_q \nonumber\\
\frac{d\bar{\alpha}_k (B)}{dB} &=& -2 \eta_k^{(2)} \bar{\beta} -2 
\sum_q \eta_{q,k} \bar{\alpha}_q
\label{bfeq}
\end{eqnarray}

The initial conditions are $\beta(B=0)=\bar{\beta}(B=0)=1/2, ~  
\alpha_k(B=0)=\bar{\alpha}_k(B=0)=0$.
During the flow towards $B \rightarrow \infty$, the operator $b$  
changes its structure into
a complicated superposition of bath operators.

\subsection{Real-time evolution in closed form}

It is now easy to formulate an exact transformation that  
yields the operator $b(t),~ t>0$, which is time-evolved
with respect to the Hamiltonian (\ref{dhomodel}). First, the operator  
$\tilde{b}$ is trivially time evolved
with respect to the Hamiltonian (\ref{hinfty}) as
$$\tilde{b}(t) \stackrel{\rm def}{=} e^{i\tilde{H}t} \tilde{b} e^{-i 
\tilde{H}t}$$
This operation endows the coefficients in (\ref{bansatz}) with
trivial phase factors

\begin{eqnarray}
\tilde{\beta}(t)          &\stackrel{\rm def}{=}& \tilde{\beta} \cos 
(\tilde{\Delta} t) -i \tilde{\bar{\beta}} \sin (\tilde{\Delta} t) 
\nonumber\\
\tilde{\bar{\beta}}(t)    &\stackrel{\rm def}{=}& \tilde{\bar{\beta}}  
\cos (\tilde{\Delta}  t )-i \tilde{\beta} \sin (\tilde{\Delta} t) 
\nonumber\\
\tilde{\alpha}_k(t)       &\stackrel{\rm def}{=}&  \tilde{\alpha}_k   
\cos( \omega_k t) - i\tilde{\bar{\alpha}}_k \sin (\omega_k t)\nonumber\\
\tilde{\bar{\alpha}}_k(t) &\stackrel{\rm def}{=}&  \tilde{\bar 
{\alpha}}_k  \cos ( \omega_k t) -i \tilde{\alpha_k} \sin (\omega_k t)
\label{initialb}
\end{eqnarray}

leading to the operator

\begin{eqnarray}
\tilde{b}(t) &=& \tilde{\beta}(t)  (b+b^\dagger)   + \tilde{\bar 
{\beta} }(t ) (b- b^\dagger) \nonumber\\
&+& \sum_k \tilde{\alpha}_k(t) ( b_k + b_k^\dagger ) \nonumber\\
&+& \sum_k \tilde{\bar{\alpha}}_k(t) ( b_k - b_k^\dagger)
\end{eqnarray}

The second step is to obtain the operator b(t) from the operator $ 
\tilde{b}(t)$ by reverting the unitary transformation $U$
\footnote{The full unitary transformation $U$ can be expressed as
a $B$-ordered exponential, $U=T_B \exp(\int_0^\infty \eta(B)dB)$.
However, this expression is only formally useful since it
cannot be evaluated without additional approximations.} used to diagonalize the Hamiltonian (\ref{dhomodel}),  
formally represented by the relation $b(t)=U^\dagger \tilde{b}(t)U$.
For this purpose, we again make an ansatz for the flow of the operator $\tilde 
{b}(t )$

\begin{eqnarray}
b(B,t) &=& \beta (B,t) (b + b^\dagger ) + \bar{\beta}(B,t) (b-b^ 
\dagger ) \nonumber\\
&+& \sum_k \bar{\alpha_k}(B,t) (b_k - b_k^\dagger )  \nonumber\\
&+& \alpha (B,t) (b_k -b_k^\dagger ),
\label{bbt}
\end{eqnarray}

where all coefficients have both real and imaginary part, since  
the initial conditions at $B= \infty$ are given by
the complex valued expressions from (\ref{initialb}). Since the  
ansatz of Eq. (\ref{bbt}) is formally identical to
that of (\ref{bansatz}), the unitary flow of $b(B,t)$ can again be  
calculated by using the flow equations (\ref{bfeq}).
The operator $b(t)$ is now obtained by integrating the flow equations  
(\ref{bfeq}) from $B= \infty$
to $B=0$, using the parameters from (\ref{bbt}) and the initial  
conditions for $B= \infty$, as given by Eq. (\ref 
{initialb}). Since all transformations
are unitary, the operator $b^\dagger(t)$ is  
readily obtained by the hermitean conjugate of Eq.
(\ref{bbt}).

All transformations used up to now do not depend on the initial state  
of the quantum system. In calculations of
time-dependent physical quantities, the operator $b(t)$ can be  
evaluated with respect to equilibrium heat baths at finite
temperature as well as arbitrary non-equilibrium ensembles.

\subsection{Analytical results}

\begin{figure}
\includegraphics[clip=true,width=7.0cm,angle=-0]{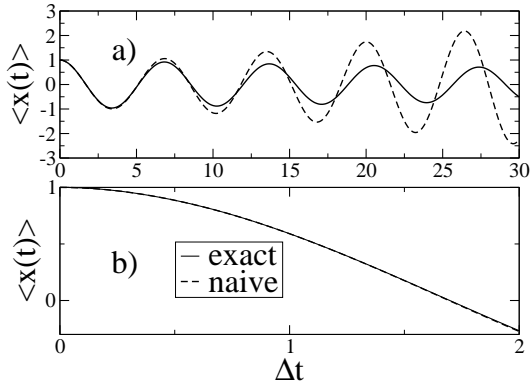}
\caption{\label{expansion} Comparison of the exact solution
Eq. (\ref{analytical}) against the naive perturbation theory of Eq.
(\ref{xtnaive}). The secular term occuring in the second order  
perturbation expansion
yields an error growing $\propto t$, see a).
In the short time limit $t\ll1$, naive perturbation theory becomes  
exact, see b).
Parameters: $a=\frac{1}{\sqrt{2}}$, $\Delta=1$;  Ohmic bath:  $J 
(\omega)=2\alpha\omega \Theta(\omega_c-\omega)$ with $\alpha=0.001$, $ 
\omega_c=10$.
}
\label{displace}
\end{figure}

{\sl 1. naive perturbation theory}\\

In analogy to classical perturbation theory as discussed in the  
introduction,
a perturbative result for quantum evolution can be obtained by  
directly expanding
the Heisenberg equation of motion for an operator. The exact time  
evolution of the
operator $\hat x = (b^\pdag+b^\dagger)/\sqrt{2}$ is

\begin{equation}
\hat{x}(t)=e^{iHt} \hat{x} e^{-iHt},
\label{heisenx}
\end{equation}

where $H$ is given by Eq. (\ref{dhomodel}). A perturbative expansion  
of Eq. (\ref{heisenx}) is typically
performed in the interaction picture, which we define by

\begin{equation}
\hat{x}_I(t)=e^{-iH_0t} \hat{x}(t) e^{iH_0t},
\label{diracx}
\end{equation}

and $H_0=\Delta b^\dagger b^\pdag +\sum_k \omega_k b_k^\dagger b_k^ 
\pdag$ is the non-interacting part of the Hamiltonian.
Eq. (\ref{diracx}) leads to the equation of motion

\begin{equation}
\frac{d \hat{x}^I(t)}{dt}=i[H_{int}^I(t),\hat{x}^I(t)]
\end{equation}

which can then be expanded as

\begin{eqnarray}
\hat{x}_I(t) &=& \hat{x} + i \int_0^t d\tau_1 [H_{int}^I(\tau_1),\hat 
{x}] \nonumber\\
              &+& i^2  \int_0^t d\tau_1 \int_0^{\tau_1} d \tau_2[H_ 
{int}^I(\tau_1),[H_{int}^I(\tau_2),\hat{x}]] \nonumber\\
              &+& O(H_{int}^{I})^3 \ .
\label{expanded}
\end{eqnarray}

Here $H_{int}^I(t)=e^{-iH_0t} \sum_k \lambda_k (b+b^\dagger ) (b_k^ 
\pdag +b_k^\dagger ) e^{iH_0t}$
is the interaction part of the Hamiltonian in the interaction  
picture. Neglecting all contributions of $O((H_{int}^{I})^3)$
or higher in the perturbative expansion will yield a result that is  
correct at least to $O(\lambda_k^3)$.

Using the inital state (\ref{coherentstate}), the first order  
contribution from Eq. (\ref{expanded}) vanishes, since
$\langle b_k^\pdag \rangle =\langle b_k^\dagger \rangle=0$.   
Likewise, all odd orders of perturbation theory vanish.
Evaluation of the two commutators needed for the second order  
contribution to $\langle \hat x(t)\rangle$ yields

\begin{eqnarray}
\langle \hat{x}(t) \rangle &=& \sqrt{2} a \cos(\Delta t) - \frac{a} 
{\sqrt{2}} \int_0^\infty d\omega \frac{4\Delta J(\omega)}{\Delta^2 -  
\omega^2}
\nonumber\\
&\times& \biggl( -\frac{2\omega}{\Delta^2-\omega^2} (\cos(\omega t) - 
\cos(\Delta t)) \nonumber\\
&+& \frac{\omega}{\Delta} \sin(\Delta t)t\biggr) + O(\lambda_k^4).
\label{xtnaive}
\end{eqnarray}

In analogy to our example of naive perturbation theory in classical  
mechanics from section \ref{feq}, a secular term
occurs in the expansion from Eq. (\ref{xtnaive}) which leads to large  
errors on long time scales (see Fig. \ref{expansion}).\\

{\sl 2. exact solution}\\

We next derive a closed analytical expression for the
quantity $\langle \hat{x}(t)\rangle$ as a benchmark for a full  
numerical solution of the flow equations.
Several other exact results for the dissipative quantum oscillator  
have been obtained by Haake and Reibold.\cite{Haake}
According to Eq. (\ref{bbt}), we have $\langle \hat{x}(t)\rangle=  
\sqrt{2} a \beta(0,t)$.
The coefficient $\beta(0,t)$ is also given by the commutator

\begin{equation}
\beta(0,t)=\frac{1}{2}[b^\pdag-b^\dagger , b^\pdag(t)]_-,
\label{commute}
\end{equation}

and using the invariance of (\ref{commute}) under the unitary flow  
(\ref{bansatz}),
we obtain

\begin{eqnarray}
\beta(0,t)&=& \frac{1}{2}[\tilde{b}^\pdag -\tilde{b}^\dagger,\tilde{b} 
(t)]\nonumber\\
&=& 2 \sum_k (\tilde{\alpha}_k^2 +\tilde{\bar{\alpha}}_k^2) \cos 
(\omega_k t).
\end{eqnarray}

It is possible to evaluate this sum in closed form by defining

\begin{eqnarray}
K(\omega)&=&\sum_k s_k^2 \delta(\omega^2-\omega_k^2) \Theta(\omega_c  
- \omega) \nonumber\\
s_k&=&2\Bigl(\frac{\omega_k}{\Delta}\Bigr)^{1/2} \tilde{\alpha}_k=2 
\Bigl(\frac{\Delta}{\omega_k}\Bigr)^{1/2} \tilde{\bar{\alpha}}_k
\end{eqnarray}

what leads to the exact result for the dynamics

\begin{equation}
\langle \hat{x}(t) \rangle=2\sqrt{2} a \int_0^\infty \omega K(\omega)  
\cos(\omega t) d\omega .
\label{analytical}
\end{equation}

For an Ohmic bath with $J(\omega)=\alpha\omega\Theta(\omega_c-\omega) 
$ , the function $K(\omega)$ can be evaluated as

\begin{eqnarray}
K(\omega)&=& (4 \alpha \omega \Delta) \biggl(16\pi^2\alpha^2  
\Delta^2\omega^2 \nonumber\\
&+& [\Delta^2 - \omega^2 +8\alpha  
\Delta (-\omega_c + \frac{\omega}{2}\textrm{ln}
(\frac{\omega + \omega_c}{\omega_c-\omega}))]^2 \biggr)^{-1} .
\end{eqnarray}

\subsection{Numerical results}

In order to illustrate the potential applications of our  
diagonalization scheme, we calculate
the expectation value $\langle \hat{x}(t)$ by a numerical integration  
of the flow equations Eq. (\ref{bfeq}). For our numerical  
calculation, we specify the spectral function of the bath as an Ohmic  
one with a sharp cutoff frequency $\omega_c$.

\begin{equation}
J(\omega)=\alpha \omega\Theta(\omega_c-\omega)
\end{equation}

This spectral function is discretized with $N=O(10^3)$ states with a  
constant energy spacing of $\Delta \omega=\frac{\omega_c}{N}$.
The systems of coupled differential equations (23) and (26) have to  
be solved separately for each point in time, since the initial  
conditions of Eq. (\ref{initialb})
depend on time.  For a finite number $N$ of bath modes, the initial  
conditions  for the bath modes have a recurrence time of $T=2\pi N /  
\omega_c$, and it is expected that the error blows up  
rapidly at this time scale. Indeed, we could
confirm this behavior numerically. Nevertheless, for times $t<T= O 
( N / \omega_c)$, a number of about $N=O(10^3)$ bath modes is  
sufficient to agree with the exact result within an error of less  
than $1\%$.\\
Summing up, numerical simulations with only $O(10^3)$ bath states  
provide excellent
agreement with analytical solutions, with finite size effects  
occuring only at time scales of $O(N)$.
In \ref{feqs}, we briefly discuss how to efficiently  
implement our transformation
scheme for finite size systems.

\begin{figure}
\includegraphics[clip=true,width=8.0cm]{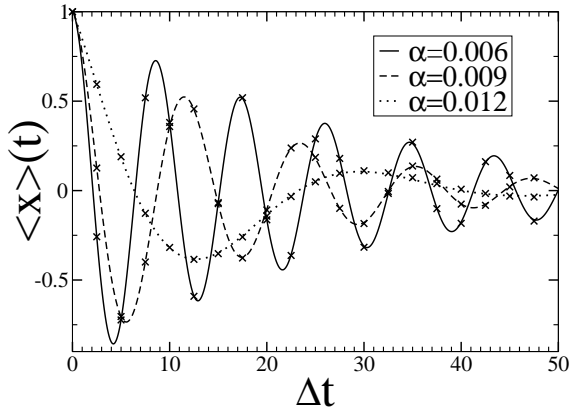}
\caption{\label{occupation} Comparison of the time-dependent  
displacement $\langle \hat{x}(t)$ of the dissipative harmonic  
oscillator,
obtained from the analytical result of Eq. (\ref{analytical})  
(continuous lines) and numerical integration of the flow equations
(crosses). Different damping strengths $\alpha$
have been chosen, and the tunneling matrix element $\Delta$ is almost  
renormalized to zero at $\alpha=0.012$, leading to
a much slower oscillation period. The curve has been normalized to 1  
by chosing $a=1/\sqrt{2}$. The comparison demonstrates that a  
numerical solution of real time observables with about 1000 bath  
states reproduces the analytical result already excellent with an  
error below $1 \%$.
}
\label{displace}
\end{figure}

\section{\label{tss}The dissipative two-level system}

The dissipative two-level system

\begin{equation}
H=-\frac{\Delta}{2} \sigma_x + \frac{\sigma_z}{2} \sum_k (b_k^\pdag +  
b_k^\dagger) + \sum \omega_k b_k^\dagger b_k^\pdag
\label{tls}
\end{equation}

is a fundamental model for the description of decoherence and  
dissipation in quantum systems.\cite{Weiss1999,Leggett1987} The two- 
state system is represented by pseudospin operators $\sigma_i, i=x,y,z 
$ and the effect of dissipation is caused by a linear coupling to a  
bosonic bath.
All bath properties can again be modeled by the spectral function $J 
(\omega)$ defined in Eq. (\ref{spectrum}).

Typically, the effect of decoherence manifests itself if the two- 
level system is prepared in an eigenstate and the
coupling to the environment is switched on subsequently. The  
observable $\langle \sigma_z(t) \rangle$ describes
then the tunneling dynamics of the initially pure quantum state.

This problem turns out to be non-trivial and nearly all known results  
rely on approximations. The approximation
scheme of the flow equation approach can be applied to this problem  
in a controlled
way as shown below. This example demonstrates the ability of our  
method to treat real-time evolution 
in non-integrable models without any problems due to secular terms. 

\subsection{Diagonalization of the Hamiltonian}

In order to approximately diagonalize the Hamiltonian (\ref{tls}), we  
employ the following generator
for the unitary flow:\cite{KehreinMielke}

\begin{eqnarray}
\eta(B) &=& i \sigma_y \sum_k \eta_k^{(y)} (b_k^\pdag +b_k^\dagger )  
+ \sigma_z \sum_k \eta_k^{(z)}(b_k^\pdag- b_k^\dagger)\nonumber\\
  &+& \sum_{kl} \eta_{kl} :(b_k^\pdag +b_k^\dagger )(b_l^\pdag-b_l^ 
\pdag ):,
  \label{gentss}
\end{eqnarray}

with $B$-dependent coefficients

\begin{eqnarray}
\eta_k^{(y)}&=&\frac{\lambda_k}{2} \Delta \frac{\omega_k-\Delta} 
{\omega_k+\Delta}, ~~\eta_k^{(z)}= -\frac{\lambda_k}{2} \omega_k \frac 
{\omega_k-\Delta}{\omega_k+\Delta},\nonumber\\
\eta_{kl}&=& \frac{\lambda_k \lambda_l \Delta \omega_l}{2(\omega_k^2- 
\omega_l^2)}
\bigl( \frac{\omega_k-\Delta}{\omega_k+\Delta} + \frac{\omega_l- 
\Delta}{\omega_l+\Delta} \bigr).
\end{eqnarray}

This generator has an important conceptional property that makes it  
different from the generator (\ref{gendho}) used
in section \ref{dho}.\footnote{The generator (\ref{gentss}) is again not 
the canonical generator in  order to leave the flowing Hamiltonian
form invariant up to higher orders in the coupling $\lambda_k$. It  
also makes use of the approximation
$\langle \sigma_x \rangle=1$ and neglects small fluctuating parts of  
this expectation value.} It does not leave the Hamiltonian  
form invariant during the flow, since it generates
additional interactions. In this case, these newly generated  
interactions are of $O(\lambda_k^3)$, which we neglect
since we consider the couplings $\lambda_k$ as a  
small expansion parameter.  For more details about the flow equation
approach to the spin-boson model we refer the reader to Ref.~\cite{Kehrein_STMP}.
Due to the expansion in the couplings $\lambda_k$, the flow equation 
calculation in equilibrium is reliable (meaning errors less than 10\%) for values $\alpha\lesssim 0.2$
for an Ohmic bath.

In the above expressions, normal ordering is
denoted by $:...:$, which ensures that the truncated higher order  
interaction terms have vanishing expectation values
with respect to the quantum state used for normal-ordering.
In equilibrium, normal-ordering is therefore performed with respect to the  
equilibrium ground state,
$b_k^\pdag b_{k^\prime}^\dagger=:b_k^\pdag b_{k^\prime}^\dagger: +  
\delta_{kk^\prime}n_B(k)$, where $n_B(k)$
is the Bose-Einstein distribution. This procedure is not ideal for  
real-time evolution of physical
observables out of a non-thermal initial state $| \psi_i \rangle$. In  
order to minimize our truncation
error, we define a more general normal-ordering procedure

\begin{eqnarray}
b_k^\pdag b_{k^\prime}^\dagger &=& :b_k^\pdag b_{k^\prime}^\dagger: +  
\delta_{kk^\prime} n_B(k)+C_{kk^\prime}\nonumber\\
C_{kk^\prime} &\stackrel{\rm def}{=}& \langle \psi_i | b_k^\pdag b_{k^ 
\prime}^\dagger | \psi_i \rangle -
\delta_{kk^\prime} n_B(k).
\label{normalorder}
\end{eqnarray}

Here, the correct non-thermal initial state is used for normal-ordering.
The flow equations for the Hamiltonian then read:

\begin{eqnarray}
\frac{d \Delta}{dB} &=& -2\sum_k \lambda_k \eta_{k^\prime}^{(y)}(1 +  
2 n_B(k)\delta_{kk^\prime} +C_{kk^\prime}) \nonumber\\
\frac{d \lambda_k}{dB} &=& -(\omega_k-\Delta)^2 \lambda_k + 2 \sum_l  
\lambda_k \eta_{kl}  \nonumber\\
\frac{d E}{dB} &=& - \sum_k \lambda_k \eta_k^{(z)}
\label{feqtls}
\end{eqnarray}

In the limit $B \rightarrow \infty$, the interaction part of the  
Hamiltonian decays completely and the
transformed Hamiltonian is given by

\begin{equation}
\tilde{H}= -\frac{\tilde{\Delta}}{2}\sigma_x + \sum_k \omega_k b_k^ 
\dagger b_k^\pdag
\label{freetls}
\end{equation}

with a renormalized tunneling matrix element $\tilde{\Delta}$.
It has been shown in Ref. \cite{KehreinMielke} that $\tilde 
{\Delta}$
obeys the correct universal scaling behavior
for Ohmic baths\cite{Leggett1987}, $\tilde{\Delta} \propto \Delta (\Delta/\omega_c )^ 
{\alpha/(1-\alpha)}$.

\subsection{Real-time evolution of operators}

The truncation scheme for the flow of the Hamiltonian (\ref{tls}) can  
be employed in the same way for
the transformation of the spin operators $\sigma_i, ~~i=x,y,z 
$. In this section, only the transformations
of the operator $\sigma_x$ are presented. Details of the  
transformations of the operators $\sigma_y$ and
$\sigma_z$ are given in appendix \ref{feqs}.

An ansatz for the flow of the operator $\sigma_x$ is formally given  
by the commutator $[\eta,\sigma_x]$,
which contains all contributions to the flow of $\sigma_x$ which are  
of first order in the
couplings $\lambda_k$. For convenience, we parametrize this ansatz  
for the flowing operator $\sigma_x(B)$
as

\begin{eqnarray}
\sigma_{x}(B) &=& h(B)\,\sigma_{x}
+\sigma_{z}\sum_{k}\left(\chi_{k}(B)\,b^{\pdag}_{k}+\bar\chi_{k}(B) 
\,b^{\dagger}_{k}\right) \nonumber \\
&+&\alpha(B)
+i\sigma_{y}\sum_{k}\left(\mu_{k}(B)\,b^{\pdag}_{k}-\bar\mu_{k}(B)\,b^ 
{\dagger}_{k}\right)
\label{sigmaxb}
\end{eqnarray}

where $\bar\mu_{k}$ and $\bar\chi_{k}$ are related to $\mu_{k}$ and $ 
\chi_{k}$ by complex conjugation.
All newly generated terms in the differential equation $\frac{d  
\sigma_x(B)}{dB}=[\eta(B),\sigma_x(B)]$
are of $O(\lambda_k^2)$ in the coupling constants. These truncated  
terms are normal-ordered according to the convention
(\ref{normalorder}), which finally determines the differential flow  
of the coupling constants as

\begin{eqnarray}
\frac{dh}{dB}&=&-\sum_{k} \left(\eta_{k}^{(y)}(\chi_{k}^{\pdag}+\bar 
\chi_{k}^{\pdag})
+\eta_{k}^{(z)}(\mu_{k}^{\pdag}+\bar\mu_{k}^{\pdag})\right) \nonumber \\
&&-4\sum_{k,l}\eta_{k}^{(y)}\,C_{kl}^{\pdag} (\chi_{l}^{\pdag}+\bar 
\chi_{l}^{\pdag}) \nonumber \\
\frac{d\chi_{k}}{dB}&=&2\,h\,\eta_{k}^{(y)}+\sum_{l}
\left(\eta^{\pdag}_{kl}(\chi_{l}+\bar\chi_{l})+ \eta^{\pdag}_{lk}(\bar 
\chi_{l}-\chi_{l})\right)
  \nonumber \\
\frac{d\mu_{k}}{dB}&=&2\,h\,\eta_{k}^{(z)}-\sum_{l}
\left(\eta^{\pdag}_{lk}(\mu_{l}+\bar\mu_{l})+ \eta^{\pdag}_{kl}(\mu_ 
{l}-\bar\mu_{l})\right) \nonumber \\
\frac{d\alpha}{dB}&=&\sum_{k} \left(\eta_{k}^{(y)}(\mu_{k}^{\pdag}+ 
\bar\mu_{k}^{\pdag})
+\eta_{k}^{(z)}(\chi_{k}^{\pdag}+\bar\chi_{k}^{\pdag})\right)
\label{feqsigmax}
\end{eqnarray}

with the initial conditions $h(B=0)=1$, $\chi_k(B=0)=\mu_k(B=0)=\alpha 
(B=0)=0$.
In the thermodynamic limit, the observable $\sigma_x$ decays  
completely, $h(B \rightarrow \infty)=0$.\cite 
{Kehrein_STMP,KehreinMielke}

For the application to real-time evolution problems, it is again  
straightforward to obtain
the time-evolved operator $\tilde{\sigma}_x(t)$ by evaluating
$\tilde{\sigma}_x(t)=e^{i\tilde{H}t} \tilde{\sigma}_x e^{-i\tilde{H}t} 
$ with the diagonal Hamiltonian $\tilde{H}$
from (\ref{freetls}). It is easy to see that the transformed  
observable (\ref{sigmaxb}) remains invariant
under time evolution, and only its coefficients change to time- 
dependent functions

\begin{eqnarray}
\tilde\chi_{k}(t)&=&\big(\tilde\chi_{k}(0)\cos(\tilde\Delta t)
+i\,\tilde\mu_{k}(0)\sin(\tilde\Delta t)\big)e^{-i\omega_{k} t}  
\nonumber\\
\tilde\mu_{k}(t)&=&\big(\tilde\mu_{k}(0)\cos(\tilde\Delta t)
+i\,\tilde\chi_{k}(0)\sin(\tilde\Delta t)\big)e^{-i\omega_{k} t} , ~~
\label{coefft}
\end{eqnarray}

where $\tilde{\alpha}$ and $\tilde{h}$ remain unchanged since these  
contributions commute with $\tilde{H}$.

The operator $\tilde{\sigma}_x(t)$ with coefficients (\ref{coefft})  
can be regarded as an effective
operator for the calculation of observables in real-time with an  
error of $O(\lambda_k^2)$.
Although this effective operator relies on a perturbative expansion  
of the flow equations,
it is able to correctly describe observables on all time scales, e.g.  
the error remains
controlled also for long times. This property can be understood
from the analogy to canonical perturbation theory of
classical mechanics, where the Hamiltonian
function is first transformed to normal form.

Using Eq. (\ref{feqsigmax}) together with the initial conditions (\ref 
{coefft}),
the effective operator $\tilde{\sigma}_x(t)$ can be integrated
back into the inital basis of the problem, thereby inducing a 
non-perturbative solution of the Heisenberg equation of motion
for the operator $\sigma_x$, as illustrated in Fig. \ref 
{forwardback}. Formally, this  solution for the operator $ 
\sigma_x(t)$
is given as

\begin{eqnarray}
\sigma_x(t)&=& h(t)\sigma_x + \sigma_z \sum_k \biggl( \chi_k(t)b_k^ 
\pdag  +\bar\chi_k b_k^\dagger \biggr) \nonumber\\
&=& \alpha(t) +i\sigma_y \sum_k \biggl( \mu_k(t)b_k^\pdag - \bar \mu_k 
(t) b_k^\dagger \biggr)
\label{effsigx}
\end{eqnarray}

After the coefficients $h(t), \chi_k(t), \bar\chi_k(t), \mu_k(t),  
\bar \mu_k(t), \alpha(t)$ have been obtained,
e.g. by a numerical solution of the flow equations, any desired  
correlation function of the operator $\sigma_x(t)$
can be calculated.

\subsection{Applications}

Although it is in principle possible to obtain analytical  
approximations to flow equations like
(\ref{feqsigmax}) and (\ref{feqtls}), we  solve the flow equations  
for the spin operators
numerically in this paper.
However, these numerical solutions rely on the truncation scheme we  
introduced above. In
order to check the accuracy of this truncation scheme, we first  
analyze it by comparing it
against an exact result.\\

{\sl 1. Comparison against exact results}\\

An exact analytical solution of the spin-boson model can be easily  
obtained at the so called pure dephasing point, where $\Delta=0$.
Since $\sigma_z$ is a conserved quantitity in this case, the  
environment can be traced out analytically.
An initial state that contains locally a pure state $| \phi \rangle= 
\frac{1}{\sqrt{2}}(| \downarrow \rangle + | \uparrow \rangle)$ will
become entangled with the environment during the course of time. A  
measure for this
process is given by the decay of the off-diagonal matrix element of  
the reduced density matrix,
$\rho_{\uparrow \downarrow }(t)= \langle \phi |  Tr_{\textrm{boson}}
\{ \rho(t)\}| \phi \rangle$, and it was shown \cite{Palma}
that this can be written as $\rho_{\uparrow \downarrow }(t)=\rho_ 
{\uparrow \downarrow } \exp(-\Gamma (t))$
with

\begin{equation}
\Gamma(t)=\frac{1}{\pi} \int_0^\infty d\omega J(\omega ) \coth \bigl 
( \frac{\omega}{2T} \bigr) \frac{1-2\cos(\omega t)}{\omega^2}.
\label{exact}
\end{equation}

\begin{figure}
\includegraphics[clip=true,width=8.0cm]{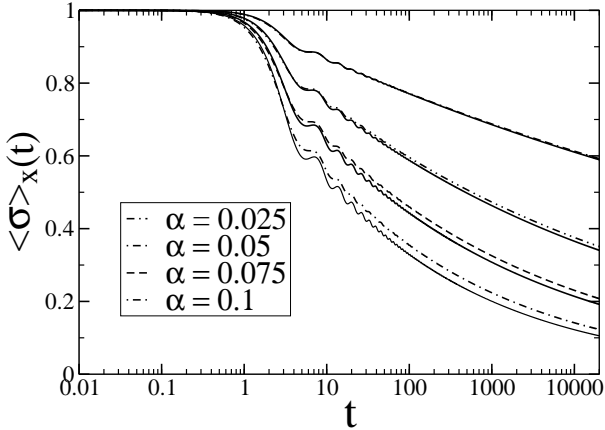}
\caption{\label{decoherence} Decay of the off-diagonal matrix element  
$\rho_{\uparrow \downarrow }(t)$ of the reduced density matrix
of the spin-boson model without tunnel splitting ($\Delta=0$) at $T=0 
$. The dashed and dotted lines represent the exact analytical result given by (\ref 
{exact}), and the full
lines the numerical solution of the flow equations. The result  
demonstrates that the truncation scheme for the flow equations
indeed is valid on all time scales, with an error that is  
systematically controlled in $O(\alpha^2)$. The bath spectral  
function has
been discretized with $N=30000$ states and cut off at $\omega_c=1$.}
\end{figure}

Note that the quantity $\rho_{\uparrow \downarrow }(t)$ is identical  
to the observable $\langle\sigma_x(t)\rangle$, which
can be directly obtained from Eq. (\ref{effsigx}). At zero  
temperature this observable shows a sluggish decay to the
ground state expectation value $\langle \sigma_x\rangle_{GS}=0$.
We now compare this exact result with the flow equation solution at  
zero temperature.
For the given  
initial state, the quantity $\rho_{\uparrow \downarrow }(t)$
is in the flow equation scheme given by the real function $\alpha(t)$ from Eq. (\ref{effsigx}).  
We evaluate this quantity for Ohmic baths with a cutoff frequency
$\omega_c$ (superohmic baths are also possible, subohmic baths have  
not yet been treated successfully within our approach). The  
calculations show that in the regime of low damping
strengths $\alpha \leq 0.1$ the agreement with the exact result is  
excellent, see Fig. \ref{decoherence}. Deviations from the exact result
are on all time scales bounded by corrections of $O(\alpha^2)$.
The numerical results also show the small transient oscillations of  
the analytical result, which oscillate with approximately the cutoff  
frequency.
This can be interpreted as a band edge effect which would vanish if a  
smooth cutoff function like $\exp(-\omega/\omega_c)$
were used for the bath spectrum.

Notice that a numerical solution of the flow equations at finite  
temperatures (not shown here) is straightforward
and also in good agreement with the exact result (\ref{exact}). \\

{\sl 2. Decay of a polarized spin}\\

\begin{figure}
\includegraphics[clip=true,width=8.0cm]{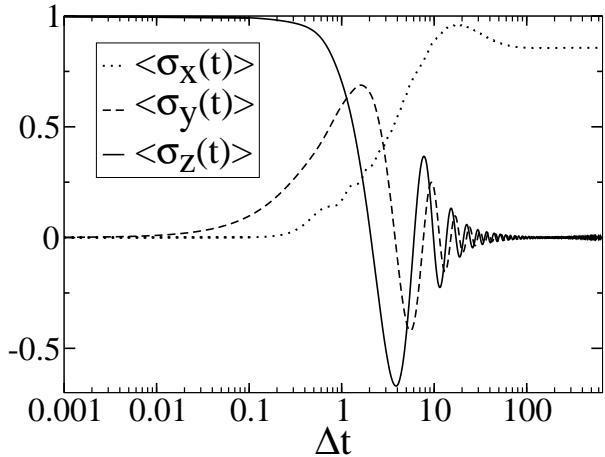}
\caption{\label{magnetization} Real time evolution of the spin  
expectation values $\langle \sigma_{x,y,z} \rangle (t)$ starting from  
a polarized
spin in z-direction with a relaxed Ohmic bath. We used a damping  
strength $\alpha=0.1$ and
the parameters $N=2000$, $\Delta=1$, $T=0$ and $\omega_c=10$.}
\end{figure}

The formulation of  
the spin-boson model was originally motivated by
the so-called quantum tunneling problem. \cite{Leggett1987,Weiss1999}  
In this problem, the two-state system is initially prepared
in an eigenstate $| \uparrow \rangle$ of the pseudospin-operator,  
corresponding to a localized state in a
double-well potential, which can be reduced to an effective two-state  
system. The bath, which we again take as Ohmic with a
frequency cutoff $\omega_c$,
is initially in thermal equilibrium and denoted by $| \textrm{bath}  
\rangle$.
Thus, the initial state $| \psi_i \rangle$ of the total system is  
given by the product state

\begin{equation}
|\psi_i \rangle = |\uparrow \rangle \otimes | \textrm{bath} \rangle.
\label{initialstate}
\end{equation}

After switching on the coupling to the dissipative environment, the  
time-dependent tunneling dynamics
between the two possible states of the two-state system is described  
by the observable

\begin{equation}
\langle \sigma_z(t) \rangle = \langle \psi_i |\sigma_z (t) | \psi_i  
\rangle,
\label{tunneling}
\end{equation}

Within the flow equation approach this is given by the coefficient 
$z(t)$ from the ansatz (\ref{flowyz}) in appendix \ref{feqs}.
We have numerically solved the flow equations (\ref{feqyz})  
corresponding to the observable (\ref{tunneling}) and depicted the  
result in
Fig. \ref{magnetization}.
As expected,  our solutions show long-time stability without secular  
terms analogous to canonical perturbation theory.
For intermediate times, our curves agree well with the 
well-established NIBA approximation, \cite{Weiss1999,Leggett1987}
and for long-time scales $t \gg \tilde{\Delta}^{-1}$ the expectation  
value $\langle \sigma_z(t) \rangle $
vanishes as expected.\\

{\sl 3. Non-equilibrium correlation functions}\\

\begin{figure}
\includegraphics[clip=true,width=8.0cm]{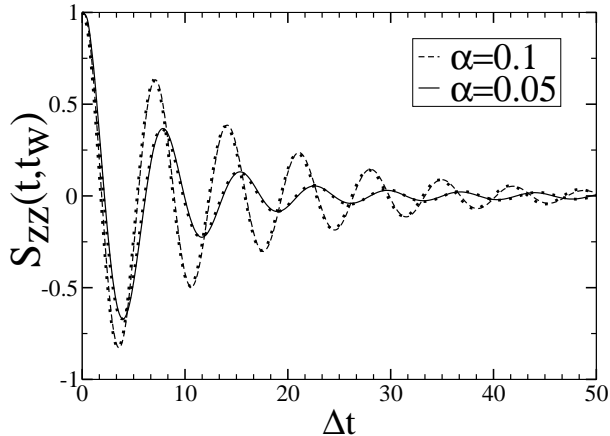}
\caption{\label{correlation} The two-time non-equilibrium correlation  
function $S_{zz}(t,t_w)$ for two different damping strengths $ 
\alpha=0.05$
and $\alpha=0.1$. The full and the dashed line show the result for $t_w=1$ and the  
dotted lines correspond to $t_w=0$. Parameters: $\Delta=1$, $ 
\omega_c=10$, $T=0$, $N=1000$.}
\end{figure}

Without any additional effort we can also calculate two-time correlation functions
based on our non-perturbative solution of the Heisenberg equations of  
motion for the spin operators. As an example we discuss the 
non-equilibrium correlation function

\begin{equation}
S_{zz}(t,t_w)=\frac{1}{2} \langle \{ \sigma_z(t+t_w),\sigma_z(t_w) \}_ 
+ \rangle,
\end{equation}

where $t_{w}$ is the waiting time for the first measurement after switching
on the dynamics. Notice that
in thermal equilibrium this correlation function is time translation  
invariant and does not depend
on the waiting time $t_w$. It is the relevant quantity for, e.g.,  
neutron scattering experiments
\cite{Weiss1999}. Experimental results for this correlator were for  
example obtained for hydrogen trapped by oxygen in niobium
\cite{Steinbinder_SZZ}, where protons can tunnel between two trap  
sites of the $Nb(OH)_x$ sample.

For studying the non-equilibrium dynamics of the spin-spin correlation function,
we again use the  quantum state $| \psi_{i} \rangle$ 
from (\ref{initialstate}) with the bath at zero temperature as the initial state.
Within the flow equation approach, $S_{zz}(t,t_{w})$ is readily evaluated as

\begin{eqnarray}
S_{zz}(t,t_w) &=& z(t+t_w)z(t_w)+ y(t+t_w)y(t_w)\nonumber\\
               &+& \sum_k [\bar{\alpha}_k (t+t_w)\bar{\alpha}_k(t_w)  
+ \alpha_k (t+t_w) \alpha_k (t_w)].\nonumber\\
\end{eqnarray}

All coefficients are explained in Appendix \ref{feqs}.
It turns out from numerical calculations (see Fig. \ref 
{correlation} ) that the dependence on $t_w$ is only weak in the  
limit of small damping strengths and the correlations are very close  
to the equilibrium behavior.

\section{\label{disc}Discussion}

Our study of real-time dynamics in dissipative quantum systems was  
motivated by
developing an analogous method to canonical perturbation theory as known from  
classical mechanics. 

Our results show:
(i) The unitary transformation scheme of the flow-equation  
approach reproduces the
well-known advantages of canonical perturbation theory from classical mechanics,
especially the absence of secular terms in the time evolution. For example in the  
spin-boson model, physical observables can be calculated reliably  
on all time scales
since perturbation theory is performed in a unitarily transformed  
basis.
(ii) The identification of a suitable expansion parameter is crucial  
to obtain reliable
results. Notice that the implementation of the flow equation scheme  
is not restricted
to bosonic baths or to impurity models, see e.g. Refs.  
\cite{Moeckel2008,Hackl2008}.
(iii) Our method is particularly suitable for studying different initial states, since
all transformations and approximations are performed on the  
operator level.

\ack
We acknowledge the financial support through SFB/TR~12 of the Deutsche Forschungsgemeinschaft. 
A.H. also  acknowledges support through SFB~608 and S.K. acknowledges support through 
the Center for Nanoscience  CeNS  Munich, and
the German Excellence Initiative via the Nanosystems Initiative
Munich  NIM .

\appendix

\section{\label{feqs} Additional flow equations for spin operators}

{\sl 1. Derivation of flow equations}\\

We briefly provide the flow equation transformations for the spin  
operators $\sigma_y$ and $\sigma_z$, that are constructed in complete analogy  
to the example from section \ref{tss}.
Using the generator (\ref{gentss}), the ans\"atze for the flowing  
spin components $\sigma_y$ and $\sigma_z$
read

\begin{eqnarray}
\sigma_y(B)=h^{(y)}(B)\sigma_y + i\sigma_x \sum_k \chi_k^{(y)}(B)  
(b_k-b_k^\dagger) +O(\lambda_k^2)\nonumber\\
\sigma_z(B)=h^{(z)}(B)\sigma_z + i\sigma_x \sum_k \chi_k^{(z)}(B) (b_k 
+b_k^\dagger)+O(\lambda_k^2)\nonumber\\
\label{ansatze}
\end{eqnarray}

The flow equations for these operators are readily derived as

\begin{eqnarray}
\frac{dh^{(z)}}{dB}&=& \sum_{kk^\prime} \eta_k^{(y)} \chi_{k^\prime}^ 
{(z)} (2 \delta_{kk^\prime} \textrm{cotanh} (\frac{\beta \omega_k}{2}) +8 C_ 
{kk^\prime} )\nonumber\\
\frac{d\chi_k^{(z)}}{dB} &=& -2 \eta_k^{(y)} h^{(z)}(B) -2 \sum_l  
\eta_{kl} \chi_l^{(z)}
\label{flowz}
\end{eqnarray}

and

\begin{eqnarray}
\frac{dh^{(y)}}{dB}&=& -\sum_{kk^\prime} \eta_k^{(z)} \chi_{k^\prime}^ 
{(y)} (2 \delta_{kk^\prime} \textrm{cotanh} (\frac{\beta \omega_k}{2})) 
\nonumber\\
\frac{d\chi_k^{(y)}}{dB} &=& -2 \eta_k^{(z)} h^{(y)}(B) -2 \sum_l  
\eta_{lk} \chi_l^{(y)}
\label{flowy}
\end{eqnarray}

Next, we solve the Heisenberg equations of motion for the operators $ 
\tilde{\sigma}_y$ and $\tilde{\sigma}_z$,
which have the formal solution $\tilde{\sigma}_{y/z}(t) =e^{it\tilde 
{H}} \tilde{\sigma}_{y/z} e^{-it\tilde{H}}$,
with the Hamiltonian $\tilde{H}$ given by Eq. (\ref{freetls}).
The solutions read:

\begin{eqnarray}
\tilde{\sigma}_{y}(t) &=& \tilde{h}^{(y)} \sigma_y \cos(\frac{\tilde 
{\Delta}}{2}t) + \tilde{h}^{(y)} \sigma_z\sin(\frac{\tilde{\Delta}}{2} 
t) \nonumber\\
&+& i\sigma_x \sum_k \tilde{\chi}_k^{(y)} (e^{-i\omega_kt}b_k-e^{i 
\omega_kt}b_k^\dagger)\nonumber\\
\tilde{\sigma}_{z}(t) &=& - \tilde{h}^{(z)} \sigma_y\sin(\frac{\tilde 
{\Delta}}{2}t) + \tilde{h}^{(z)} \sigma_z \cos(\frac{\tilde{\Delta}} 
{2}t) \nonumber\\
&+&i\sigma_x \sum_k \tilde{\chi}_k^{(z)} (e^{-i\omega_kt}b_k+e^{i 
\omega_kt}b_k^\dagger).
\end{eqnarray}

These operators differ formally from those of Eq. (\ref{ansatze}),  
therefore we chose a different ansatz for the transformation
of these operators,

\begin{eqnarray}
\sigma_{y/z}(B,t) &=& \sigma_x \sum_k (i\alpha_k(B,t)(b_k^\pdag -b_k^ 
\dagger) \nonumber\\
&+&\bar{\alpha}_k(B,t)(b_k^\pdag +b_k^\dagger)\nonumber\\
&+& y(B,t) \sigma_y +  z(B,t) \sigma_z + O(\lambda_k^2),
\label{flowyz}
\end{eqnarray}

which is for both $\sigma_y$ and $\sigma_z$ identical.
This ansatz yields the time-dependent flow equations

\begin{eqnarray}
\frac{d\alpha_k(B,t)}{dB}&=& -2 y(B,t) \eta_k^{(z)} -2 \sum_l \alpha_l 
(B,t) \eta_{lk} \nonumber\\
\frac{d\bar{\alpha_k}(B,t)}{dB} &=& -2 z(B,t) \eta_k^{(y)} + 2 \sum_l  
\bar{\alpha}_l(B,t) \eta_{kl} \nonumber\\
\frac{d z(B,t)}{dB} &=& 2 \sum_k  \bar{\alpha} (B,t) \eta_k^{(y)} + 8  
\sum_{kk^\prime} \eta_k^{(y)}
\bar{\alpha}_{k^\prime}^{(z)} C_{kk^\prime}\nonumber\\
\frac{d y(B,t)}{dB} &=& 2 \sum_k  \alpha_k (B,t) \eta_k^{(z)} - 8  
\sum_{kk^\prime} \eta_k^{(z)}
\bar{\alpha}_{k^\prime}^{(y)} C_{kk^\prime},\nonumber\\
\label{feqyz}
\end{eqnarray}

with the initial conditions

\begin{eqnarray}
\tilde{\alpha}_k(t) &=& \cos(\omega_k t) \tilde{\chi}_k^{(y)}\nonumber\\
\tilde{\bar{\alpha}}_k(t) &=& \sin(\omega_k t) \tilde{\chi}_k^{(y)} 
\nonumber\\
\tilde{y}(t) &=& \tilde{h}^{(y)}\cos(\frac{\tilde{\Delta}}{2}t) 
\nonumber\\
\tilde{z}(t) &=& \tilde{h}^{(y)}\sin(\frac{\tilde{\Delta}}{2}t)
\label{initials}
\end{eqnarray}

for $\tilde{\sigma}_{y}(t)$ and

\begin{eqnarray}
\tilde{\alpha}_k(t) &=& -\sin(\omega_k t) \tilde{\chi}_k^{(z)} 
\nonumber\\
\tilde{\bar{\alpha}}_k(t) &=& \cos(\omega_k t) \tilde{\chi}_k^{(z)} 
\nonumber\\
\tilde{y}(t) &=& -\tilde{h}^{(z)}\sin(\frac{\tilde{\Delta}}{2}t) 
\nonumber\\
\tilde{z}(t) &=& \tilde{h}^{(z)}\cos(\frac{\tilde{\Delta}}{2}t)
\end{eqnarray}

for  $\tilde{\sigma}_{z}(t)$. The effective solutions $\sigma_{y/z}(t) 
$ of the respective Heisenberg equations of motion
are now obtained by determining Eq. (\ref{flowyz}) in the case $B=0$.\\

{\sl 2. Numerical implementation of flow equations}\\

In order to numerically integrate a set of $O(10^3)\times O(10^3)$  
coupled differential
equations, an adaptive step size fourth order Runge Kutta algorithm  
is a fast
and accurate choice. Discretizing a bosonic or fermionic bath with  
equal energy spacing and
using about $N=O(10^3)$ bath states yields very accurate solutions up  
to time
scales of $O(N/\omega_c)$.
In order to determine initial conditions like Eq. (\ref{initials}),
it is sufficient to integrate the flow equations up to $B=O(N^2/ 
\omega_c^2)$, where
only a fraction of $O(1/N)$ of the couplings $\lambda_k$ has not  
decayed exponentially yet.
In order to integrate flow equations with inital conditions like Eq.  
(\ref{initials}),
a standard Runge Kutta algorithm works not properly for large values  
of the parameter $B$,
since the exponential smallness of the couplings exceeds floating  
point precision.
In our implementation, we stored therefore the flow of the couplings  
from the
diagonalization of the Hamiltonian and supplemented it to the  
integration
of the flow equation with time-dependent inital conditions. Although  
this procedure cannot use an adaptive stepsize in order to
control the error during integration, it turned out to be very  
precise in all cases we used it.

\end{document}